\documentclass[10pt, pre, eps, twocolumn,showpacs,nofootinbib,longbibliography]{revtex4-1}
\usepackage{url,ulem}
\usepackage{fourier}
\usepackage{amssymb,amsmath}
\usepackage[dvipdfmx]{graphicx}
\usepackage{graphicx,color}

\begin{document}
\title{Low-frequency discrete breathers in long-range systems without on-site potential}
\author{Yoshiyuki Y. Yamaguchi}
\email{yyama@amp.i.kyoto-u.ac.jp}
\affiliation{
  Department of Applied Mathematics and Physics, 
  Graduate School of Informatics, Kyoto University, 
  Kyoto 606-8501, Japan}
\author{Yusuke Doi}
\email{doi@ams.eng.osaka-u.ac.jp}
\affiliation{
  Department of Adaptive Machine Systems,
  Graduate School of Engineering, Osaka University,
  2-1 Yamadaoka, Suita, Osaka 565-0871, Japan}

\begin{abstract}
  We propose a new mechanism of long-range coupling
  to excite low-frequency discrete breathers without the on-site potential.
  This mechanism is universal in long-range systems
  irrespective of the spatial boundary conditions,
  of topology of the inner degree of freedom,
  and of precise forms of the coupling functions.
  The limit of large population is theoretically discussed
  for the periodic boundary condition.
  Existence of discrete breathers is numerically demonstrated
  with stability analysis.
\end{abstract}

\maketitle

{\it Introduction:}
The discrete breather (DB) or the intrinsic localized mode 
are spatially localized and temporally periodic solutions
in spatially discrete systems
\cite{Sievers1988,Campbell2004,Flach2008,Yoshimura2015}.
An important difference from solitons is that DBs are more ubiquitous
since DBs do not require neither integrability of systems
nor the balance between dispersion and nonlinearity
but require solely spatial discreteness and nonlinearity.
The discreteness suppresses the linear band of eigenfrequencies
in a bounded region, and nonlinearity provides a frequency
outside of the band. Thus, the localized oscillation 
can be maintained without accepting dispersion by the linear modes.
From its ubiquitousness,
it is expected that DBs can be excited in
various practical systems such as crystals\cite{Dmitriev2015,Bajars2015}, spin
lattices\cite{Lai1997,Schwarz1999,Zolotaryuk2001,Tang2014,Lakshmanan2014},
electrical lattices\cite{English2012,English2013}, and mechanical
systems\cite{Sato2003,Kimura2009}.

The spatial discreteness is modeled by a lattice system,
where each lattice point has a nonlinear oscillator coupling
with other oscillators through the two-body coupling potential.
The nonlinearity is classified into ``hard springs''
and ``soft springs'', where hard (resp. soft) springs have
higher (resp. lower) frequency for larger amplitude.
Correspondingly, possible frequencies of DBs are
above the band with hard springs,
or below the band with soft springs.
Making DBs with high frequency is always possible as the linear band is
upper-bounded in the discrete systems.
On the other hand, we need to add on-site potentials
to open a band gap below in systems with nearest neighbor couplings.

Lower frequency is desirable to make DBs easily.
Indeed, choice of suitable frequency of driver and systems is
important for observation of DBs\cite{Sato2003}. Choosing the suitable
driver frequency in region of lower frequency is easily accomplished
than that in region of higher frequency. Moreover, vibration with
lower frequency can be detected more easily and accurately in experiments.
It is, therefore, an important progress if we can provide a band gap below
{\it without} the on-site potential, since a new mechanism
extends diversity to design micro electro mechanical systems (MEMS),
for instance, by using the nonlinear dynamical theory.

To tackle this problem, we introduce long-range couplings
instead of frequently discussed nearest neighbor couplings.
Several studies on DBs have been performed in lattices with
long-range couplings \cite{Flach1998,Cuevas2002,Miloshevich2017},
but existence of a band gap below has not been pointed out. 
Another direction of long-range couplings in DBs
is discussed to give the pairwise interaction symmetric lattice (PISL),
which makes it easy to realize traveling DBs \cite{Doi2016}.
The coupling strength of PISL decreases as square of inter-site distance,
but the long-range couplings in this article have longer interaction 
range than PISL.

At first sight, it might be strange that the long-range couplings
help to make localized modes.
We will explain the basic idea of the mechanism
from a simple example of the mean-field coupling.
Then, we numerically show existence of the band gap below
by computing the linear eigenfrequencies, and demonstrate existence of DBs.
Moreover, we theoretically prove that the gap survives
even in the limit of large population
in the case of periodic boundary condition.

{\it Band gap below in long-range systems:}
Let us start from proving the basic idea to have the band gap below
without the on-site potential.
We consider a family of
one-dimensional lattice systems with long-range couplings
described by the Hamiltonian
\begin{equation}
  H(q,p) = \sum_{j=0}^{N-1} \dfrac{p_{j}^{2}}{2}
  + \dfrac{1}{2N_{\ast}} \sum_{j=0}^{N-1} \sum_{k=0}^{N-1} J_{jk} \phi(q_{j}-q_{k})
  + \sum_{j=0}^{N-1} U(q_{j}).
  \label{eq:Hamiltonian}
\end{equation}
The last term is the on-site potential,
which is added to demonstrate the modification of the band structure
but will be zero in our study.
Let both $\phi(q)$ and $U(q)$ be even and stable around $q=0$,
which implies $\phi'(0)=U'(0)=0$ and $\phi''(0),U''(0)>0$.
The coupling constant $J_{jk}$ depends on the site distance
between the sites $j$ and $k$, which are
$J_{jk}=1~(k=j\pm 1)$ and $0~(\text{otherwise})$
for the nearest neighbor coupling
and $J_{jk}=1~(j,k=0,\cdots,N-1)$ for the mean-field coupling.
The prefactor $N_{\ast}$ is defined by
\begin{equation}
  N_{\ast} = \dfrac{1}{N} \sum_{j=0}^{N-1} \sum_{k=0}^{N-1} J_{jk}
\end{equation}
depends on $N$ in general to ensure the extensivity of the potential term.
For instance, the mean-field coupling gives $N_{\ast}=N$.

The equations of motion are
\begin{equation}
  \ddot{q}_{j} = - \dfrac{1}{N_{\ast}} \sum_{k=0}^{N-1} J_{jk} \phi'(q_{j}-q_{k})
  - U'(q_{j})
  ~(j=0,\cdots,N-1).
\end{equation}
We assumed $\phi'(0)=U'(0)=0$, and hence
$q_{j}=0~(j=0,\cdots,N-1)$ is a fixed point.
Linearizing the above equations of motion around this fixed point,
we obtain
\begin{equation}
  \ddot{\xi}_{j} = - \sum_{k=0}^{N-1} B_{jk} \xi_{k},
  \quad
  B_{jk} = \phi''(0) A_{jk} + U''(0) \delta_{jk}
\end{equation}
where $\xi_{j}$ is the infinitesimal displacement from the fixed point,
and the matrix $A=(A_{jk})_{j,k=0,\cdots,N-1}$ is defined by
$A_{jk}:=\delta_{jk}-J_{jk}/N_{\ast}$.
We are interested in the eigenvalues of the matrix $B=(B_{jk})_{j,k=0,\cdots,N-1}$
and we focus on the problem if the eigenvalues have a gap below the band.
It is easy to find that the on-site potential uniformly raises
the eigenvalues of the matrix $B$ due to $U''(0)>0$
and hence the gap opens below the band.
We show that the long-range nature permits to have the band gap below
even the on-site potential is absent.
Hereafter, we set the on-site potential zero, $U=0$,
and consider the eigenvalues of the matrix $A$ instead of $B$.

The idea to have a gap below in long-range systems is as follows.
One of typical long-range couplings is the mean-field coupling,
and it gives the matrix $A$ as
\begin{equation}
  A_{\rm MF} = E_{N} - \dfrac{1}{N}
  \begin{pmatrix}
    1 & 1 & \cdots \\
    1 & 1 & \cdots \\
    \vdots & \vdots & \ddots \\
  \end{pmatrix}
\end{equation}
where $E_{N}$ is the identity matrix of size $N$.
The second term of $A_{\rm MF}$ has the rank $N-1$,
and hence all the eigenvalues of $A_{\rm MF}$ are $1$ except for one $0$
corresponding to the momentum conservation.
Therefore, the band of $A_{\rm MF}$ has a clear gap between $0$ and $1$.

In order to show that this band gap below survives
even if the coupling is not the mean-field but long-range,
we assume $J_{jk}=1/r_{jk}^{\alpha}~(\alpha\geq 0)$,
where $r_{jk}$ is the distance between $j$th and $k$th sites.
The eigenvalues of $A$ is determined by the boundary condition,
and we consider the periodic boundary condition
and the fixed boundary condition.
In the latter condition, we fix the $-1$th and $N$th particles,
which have interactions among all the other particles
following the rule of coupling constant $J_{jk}$.
We remark that the fixed particles break the momentum conservation
and the lowest eigenvalue of $A$ may not be zero.
To avoid divergence of $J_{jk}$ for $j=k$,
we have to redefine the values $J_{kk}~(k=0,\cdots,N-1)$.
We set $J_{kk}=1$ in the periodic boundary condition
and $J_{kk}=0$ in the fixed boundary condition
for showing that this choice is not crucial.

\begin{figure}
  \centering
  \includegraphics[width=7cm]{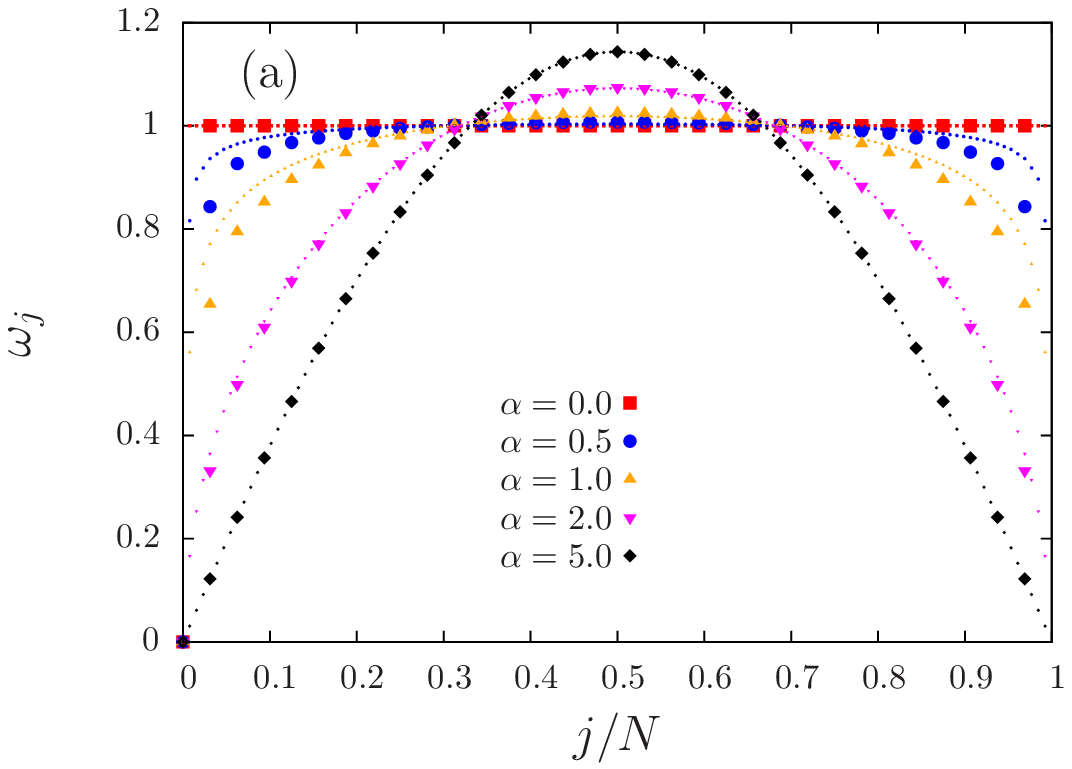}
  \includegraphics[width=7cm]{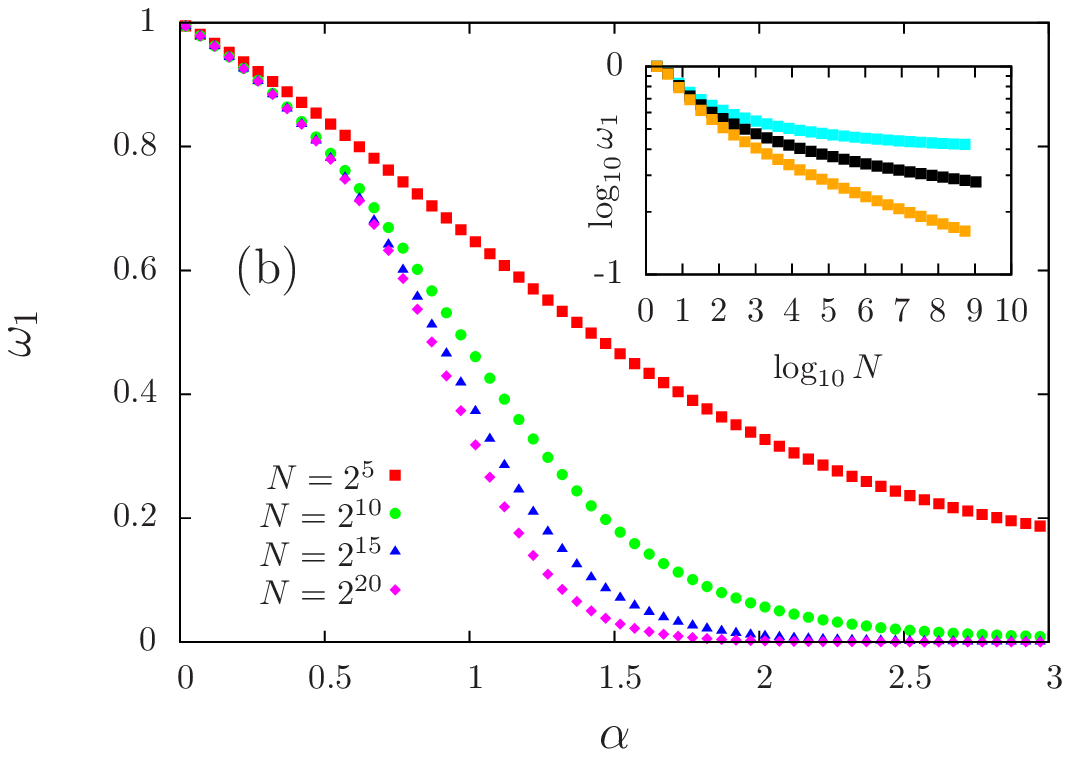}
  \caption{(color online) 
    Scaled eigenfrequencies of the linearized equations of motion,
    which are square roots of the eigenvalues of the matrix $A$.
    Periodic boundary condition.
    (a) Band structure for some values of $\alpha$.
    $N=32$ (large symbols) and $N=128$ (small symbols).
    $\omega_{0}=0$ from the total momentum conservation. 
    (b) $\alpha$ dependence of the smallest positive frequency $\omega_{1}$
    for some values of $N$.
    (inset) $N$ dependence for $\alpha=0.9,~1$ and $1.1$
    from the top to the bottom.}
  \label{fig:EigenValuesPeriodic}
\end{figure}

\begin{figure}
  \centering
  \includegraphics[width=7cm]{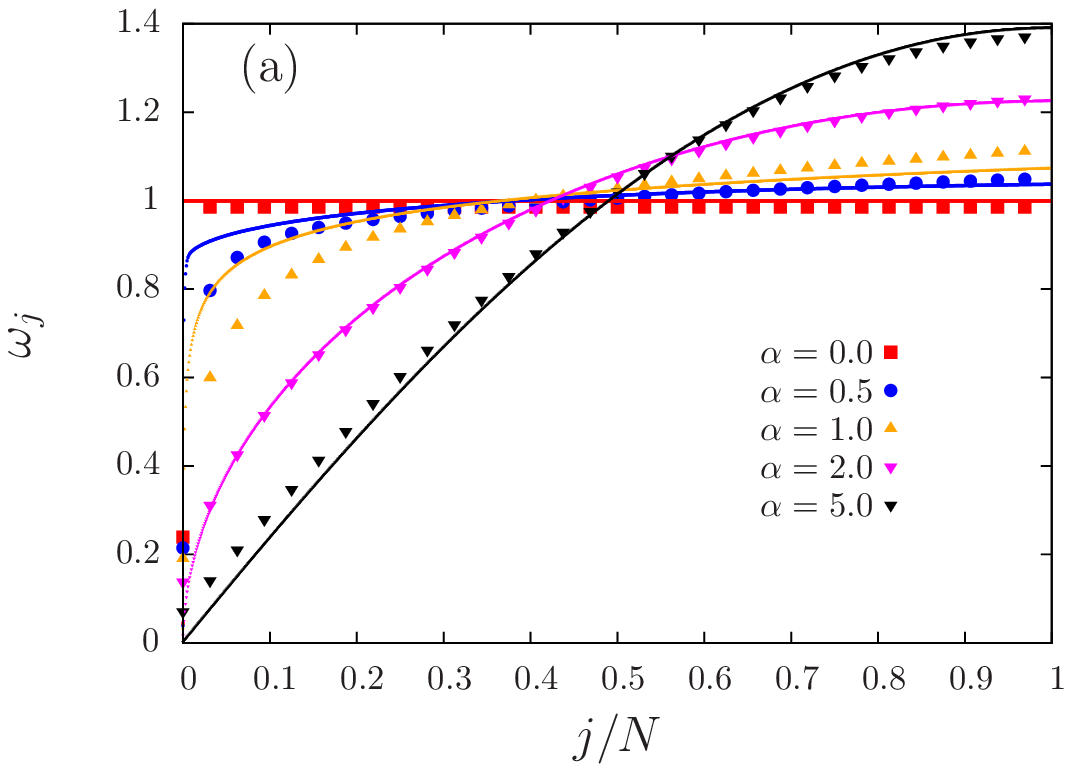}
  \includegraphics[width=7cm]{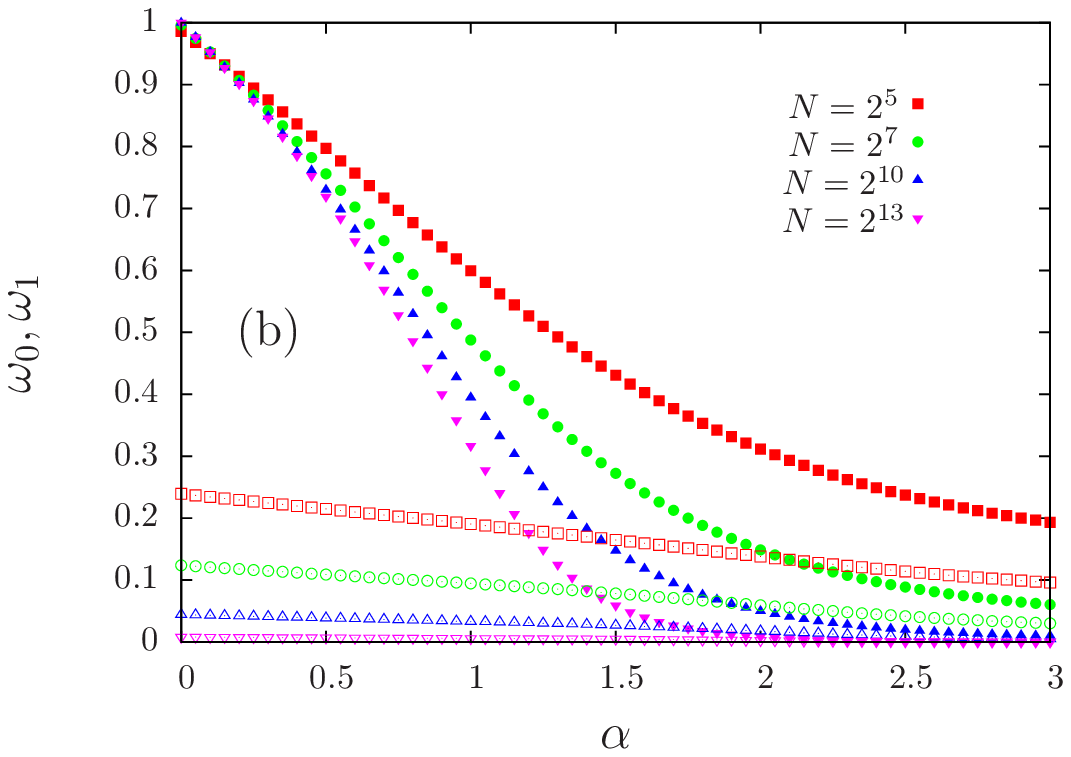}
  \caption{(color online) 
    Same with Fig.\ref{fig:EigenValuesPeriodic}
    but for the fixed boundary condition.
    See the text for the precise boundary condition.}
  \label{fig:EigenValuesFixed}
\end{figure}

Note that $\alpha\leq 1$ and $\alpha>1$ imply long-range and short-range
interaction respectively
\cite{Campa2009}, and $\alpha=2$ for PISL.
We can find the band gap below for the long-range case
in the periodic (Fig.\ref{fig:EigenValuesPeriodic})
and fixed (Fig.\ref{fig:EigenValuesFixed}) boundary conditions.
Importance of the long-range nature is confirmed
from $\alpha$-dependence of $\omega_{0}$ and $\omega_{1}$
in Figs.\ref{fig:EigenValuesPeriodic}(b) and \ref{fig:EigenValuesFixed}(b),
since the gap survives for $\alpha<1$
but it tends to vanish for $\alpha>1$ as $N$ increases.

{\it Band analysis for periodic boundary condition:}
For the periodic boundary condition,
the existence of band gap below in the limit $N\to\infty$
can be proven.
Let $N$ be even. The matrix $A$ is a circulant matrix,
and its eigenvalues are written as
\begin{equation}
  \lambda_{j} = 1 - \dfrac{1}{N_{\ast}} \left[
    J_{0} + 2 \sum_{k=1}^{N/2-1} J_{k} \cos \dfrac{2\pi jk}{N} + (-1)^{j} J_{N/2}
  \right],
\end{equation}
where $J_{|j-k|}=J_{jk}$.
Due to the periodic boundary condition,
the $0$th eigenvalue is always zero, $\lambda_{0}=0$,
and $\lambda_{N-j}=\lambda_{j}$ holds for $j=1,\cdots,N-1$.
Accordingly, the zone boundary eigenvalues are $\lambda_{1}=\lambda_{N-1}$.
Our task is to prove $\lambda_{1}=\lambda_{N-1}>0$
for the long-range case $\alpha<1$ in the limit $N\to\infty$,
which brings
\begin{equation}
  \label{eq:lambda1}
  \lambda_{1} = \lambda_{N-1} = 1 - \dfrac{2}{\kappa_{\alpha}}
  \int_{0}^{1/2} \dfrac{\cos(2\pi x)}{x^{\alpha}} dx,
  \quad
  \kappa_{\alpha} = 2 \int_{0}^{1/2} \dfrac{dx}{x^{\alpha}},
\end{equation}
where the variable $x$ is defined as $x=j/N$.
Due to $|\cos(2\pi x)|<1$ for $x\in (0,1/2)$,
we have $\lambda_{1}=\lambda_{N-1}>0$.
We underline that the above analysis is valid for $\alpha<1$
since the two integrals appearing in \eqref{eq:lambda1} diverge
from the contribution around $x=0$ for $\alpha>1$.
In the short-range case $\alpha>1$,
we have to be careful for the factor $J_{0}$, but roughly speaking, 
the main contribution to the divergence around $x=0$
may permit the approximation $\cos(2\pi x)\simeq 1$,
and $\lambda_{1}$ goes to $0$.
Indeed, this discussion is supported by the numerical computations
reported in Fig.\ref{fig:EigenValuesPeriodic}.
We, therefore, conclude that the long-range nature is important
to ensure the band gap below.
In this article, we skip the boundary case, $\alpha=1$.
Note that DBs in the long-range systems like (\ref{eq:Hamiltonian})
has been discussed in \cite{Flach1998} and \cite{Miloshevich2017}.
However the case that $\alpha<1$ has not been considered
($\alpha>1$ in \cite{Flach1998} and $\alpha=2$ in \cite{Miloshevich2017}).

{\it Examples:}
We demonstrate existence of DBs with low frequencies
in two systems having the periodic and fixed boundary conditions respectively.
The coupling constant $J_{jk}$ is fixed
as $J_{jk}=1/r_{jk}^{\alpha}~(\alpha\geq 0)$
and $J_{kk}=1$ (resp. $0$) for the periodic (resp. fixed)
boundary condition as above.
To hit the band gap below, the interaction should be ``soft springs'':
the larger amplitude gives the smaller frequency.
The first model is the so-called $\alpha$-Hamiltonian mean-field ($\alpha$-HMF)
model, which has the two-body interaction potential of
$\phi_{\rm HMF}(q) = 1 - \cos q$, where $1$ is added to adjust
the potential minimum as $0$.
The second model is a modified Fermi-Pasta-Ulam (FPU) model
whose two-body interaction potential is given by
$\phi_{\rm FPU}(q) = q^{2}/2 + \beta q^{4}/4 + \gamma q^{6}/6$.
From an analogy of the $\alpha$-HMF model
and the conventional naming of the FPU models,
we call this model as $\alpha$-FPU-$\beta\gamma$ model.
To realize the ``soft springs'', the coefficients are fixed
as $\beta=-2$ and $\gamma=1$,
which give plateaus around $x=\pm 1$.
We give the periodic boundary condition for the $\alpha$-HMF model,
and the fixed boundary condition for the $\alpha$-FPU-$\beta\gamma$ model.
Moreover, the two models demonstrate that the essential mechanism
does not depend on topology of the inner degree of freedom.
Indeed, $q$ is on the unit circle $S^{1}$ in the $\alpha$-HMF model,
and is on the real axis $\mathbb{R}$ in the $\alpha$-FPU-$\beta\gamma$ model.

\begin{figure}
  \centering
  \includegraphics[width=3.5cm]{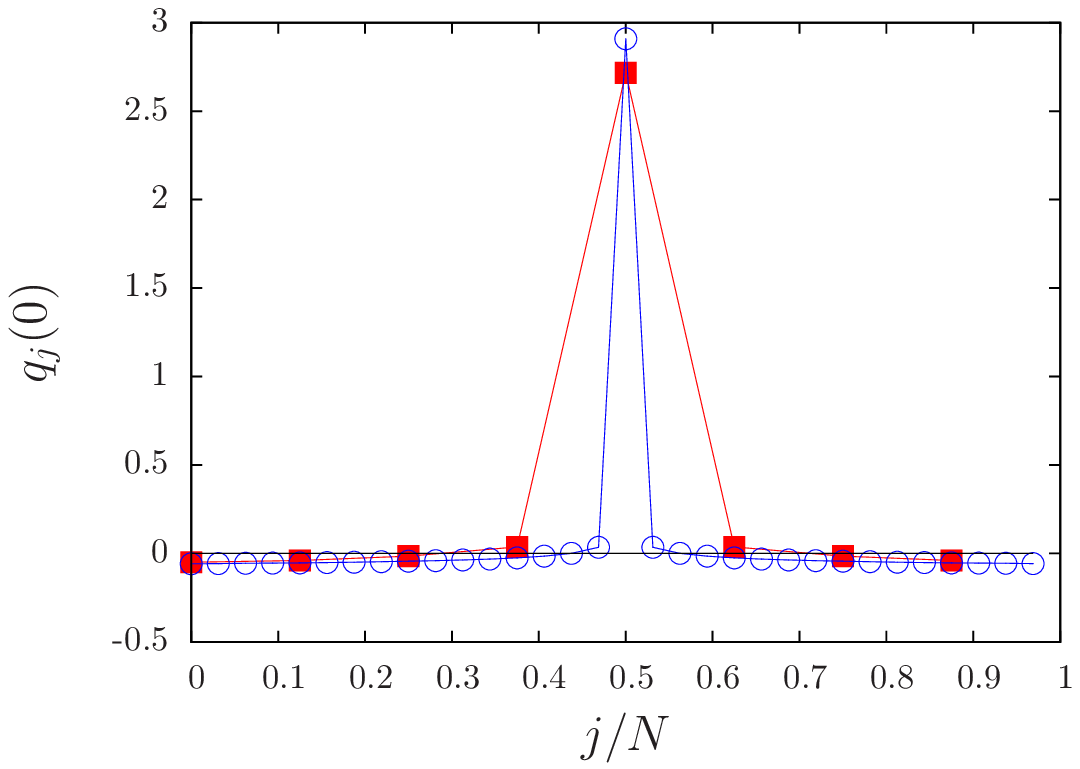}
  \includegraphics[width=3.5cm]{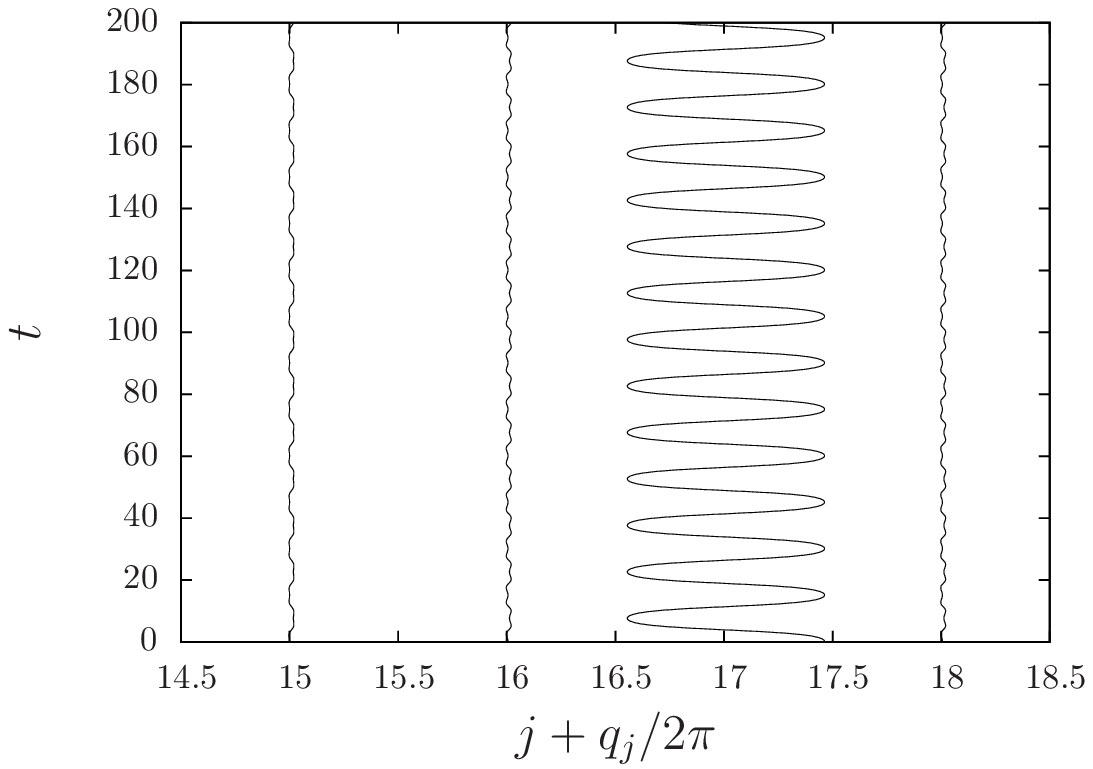}
  \includegraphics[width=3.5cm]{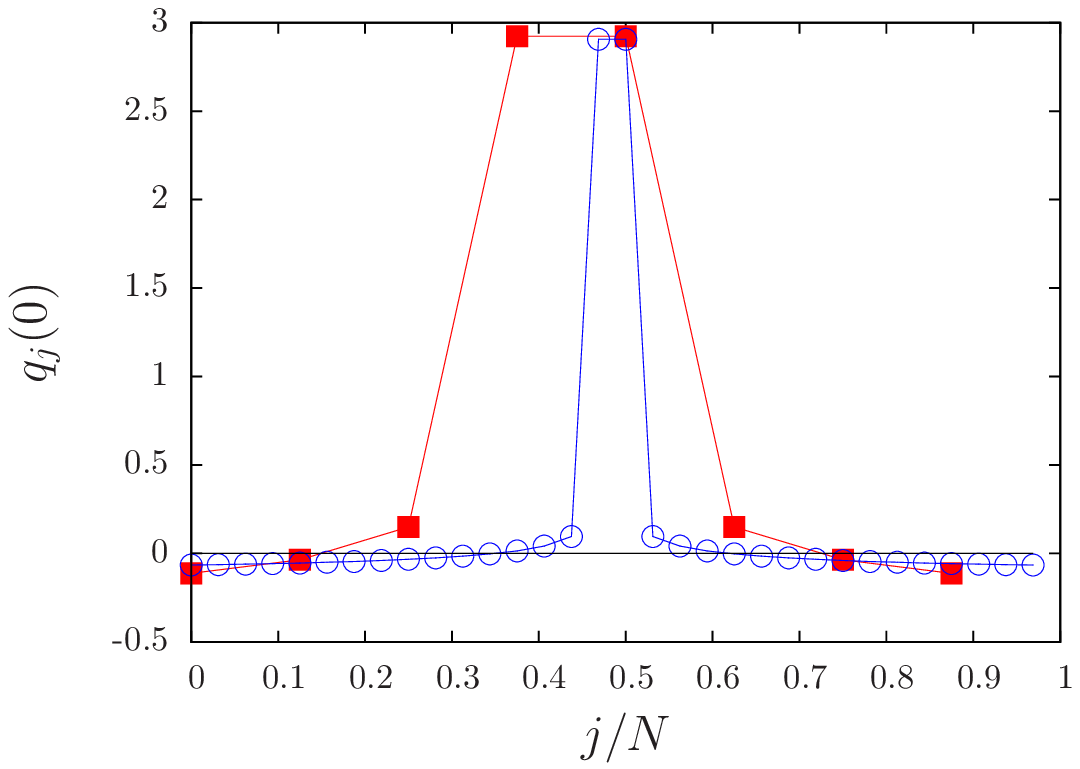}
  \includegraphics[width=3.5cm]{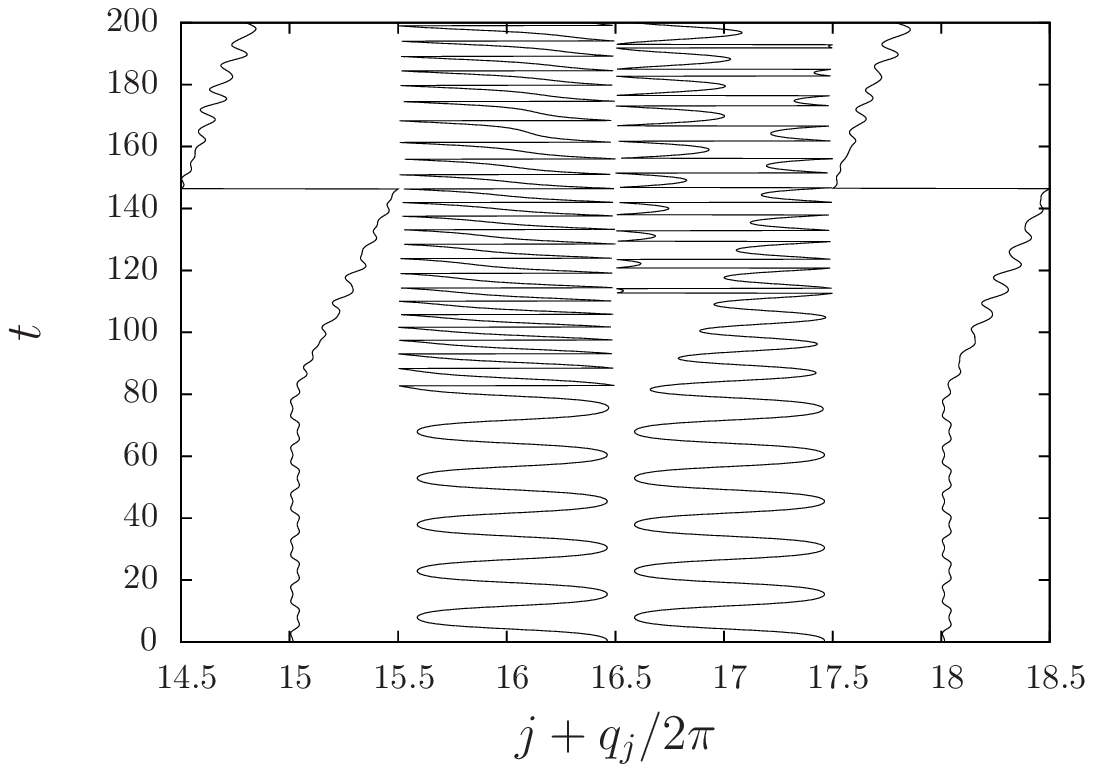}
  \includegraphics[width=3.5cm]{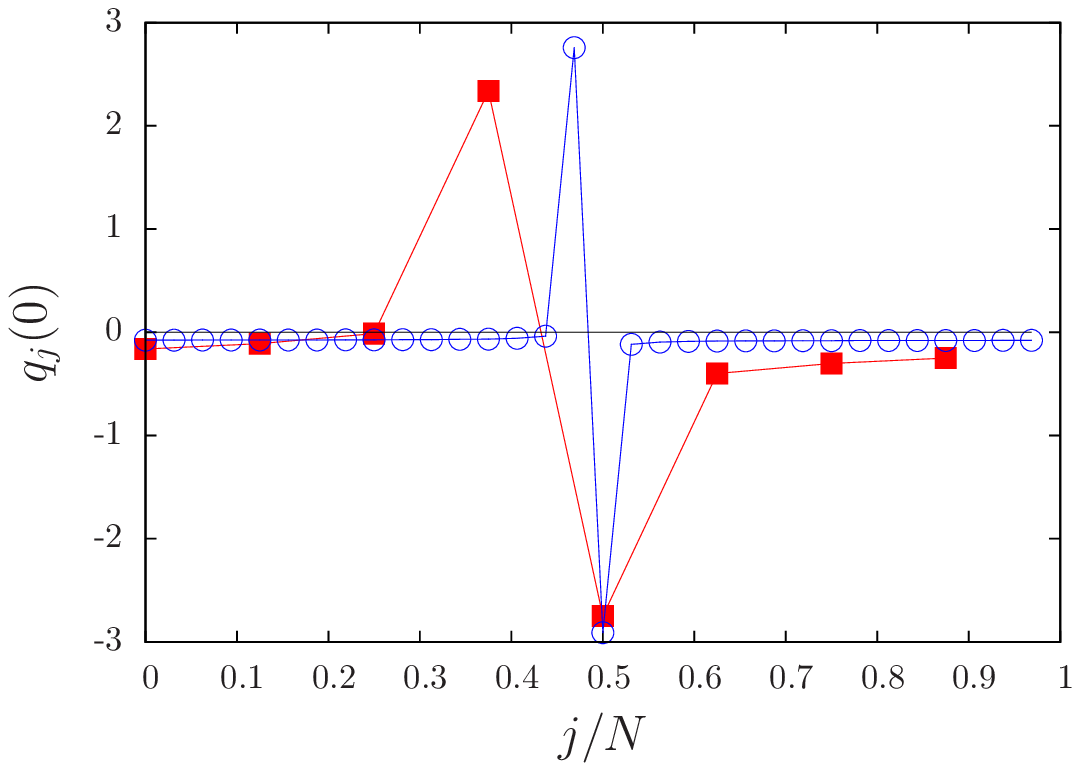}
  \includegraphics[width=3.5cm]{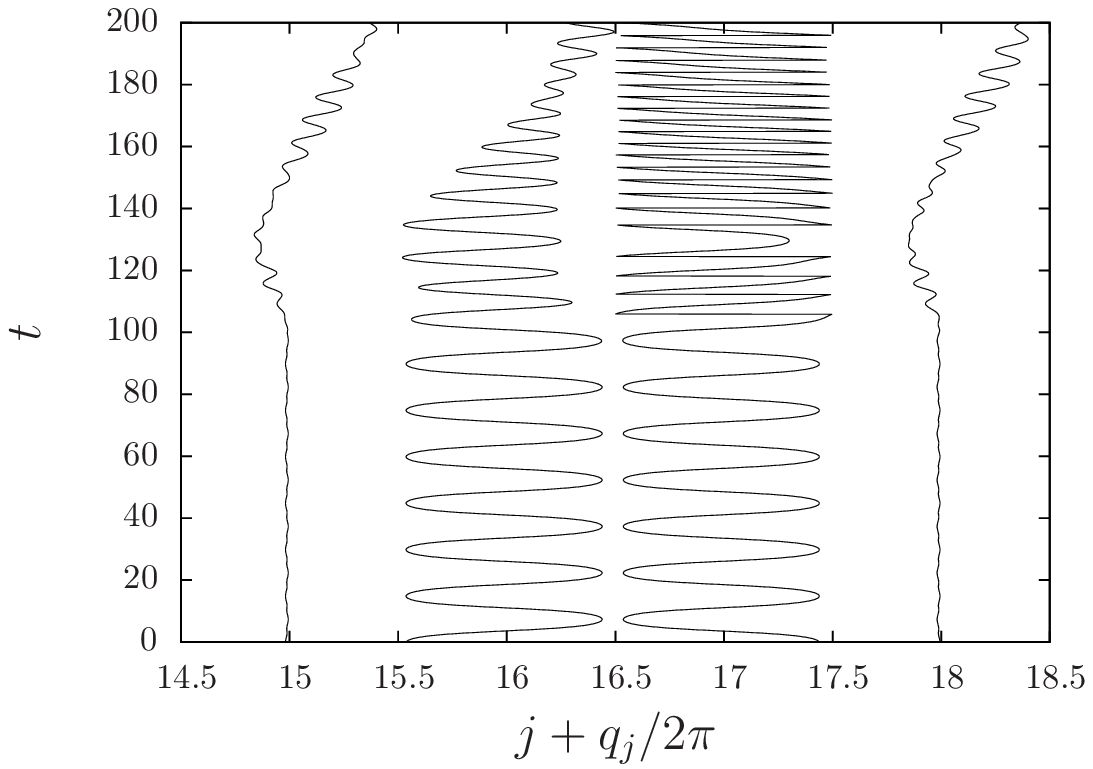}
  \caption{(color online)
    Spatially-localized periodic solutions in the $\alpha$-HMF model
    with the periodic boundary condition.
    $\alpha=0.5$.
    ST, P-in-phase, P-anti-phase modes from the top to the bottom.
    The period is $T=15$ in all the panels.
    (left) Initial configurations. $N=8$ (red filled squares)
    and $32$ (blue open circles).
    (right) Corresponding temporal evolutions of
    $q_{14},\cdots,q_{17}$ for $N=32$,
    where the amplitudes are divided by $2\pi$ for a graphical reason.}
  \label{fig:HMF}
\end{figure}

\begin{figure}
  \centering
  \includegraphics[width=3.5cm]{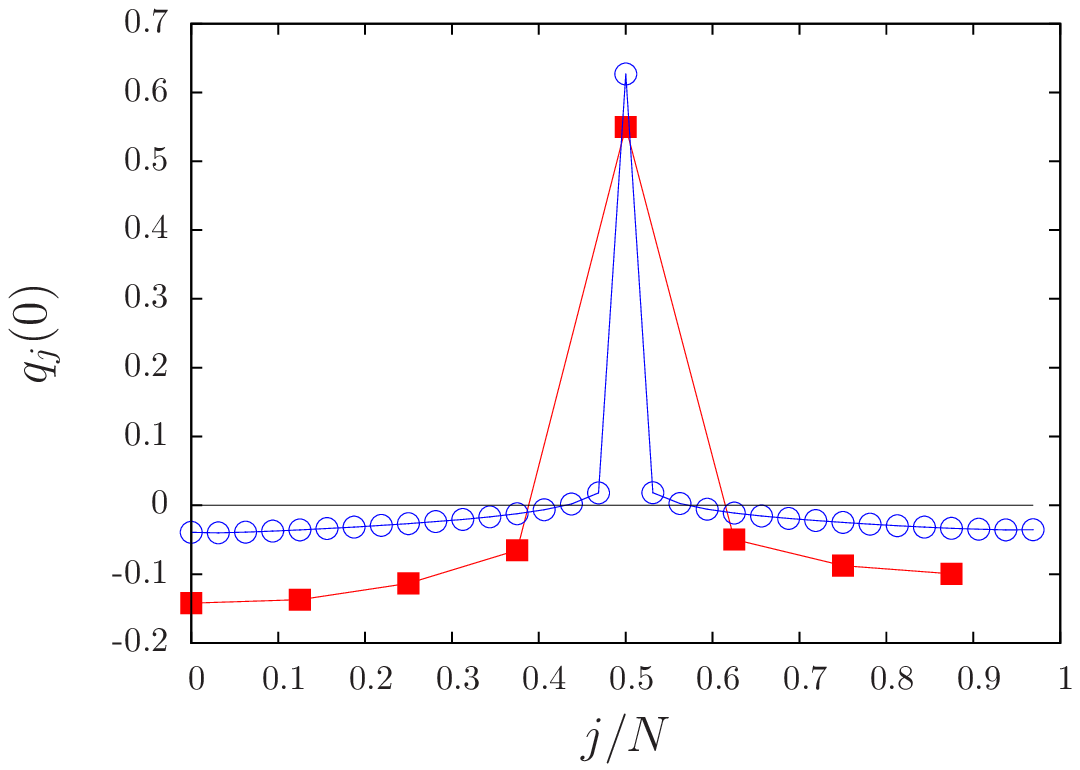}
  \includegraphics[width=3.5cm]{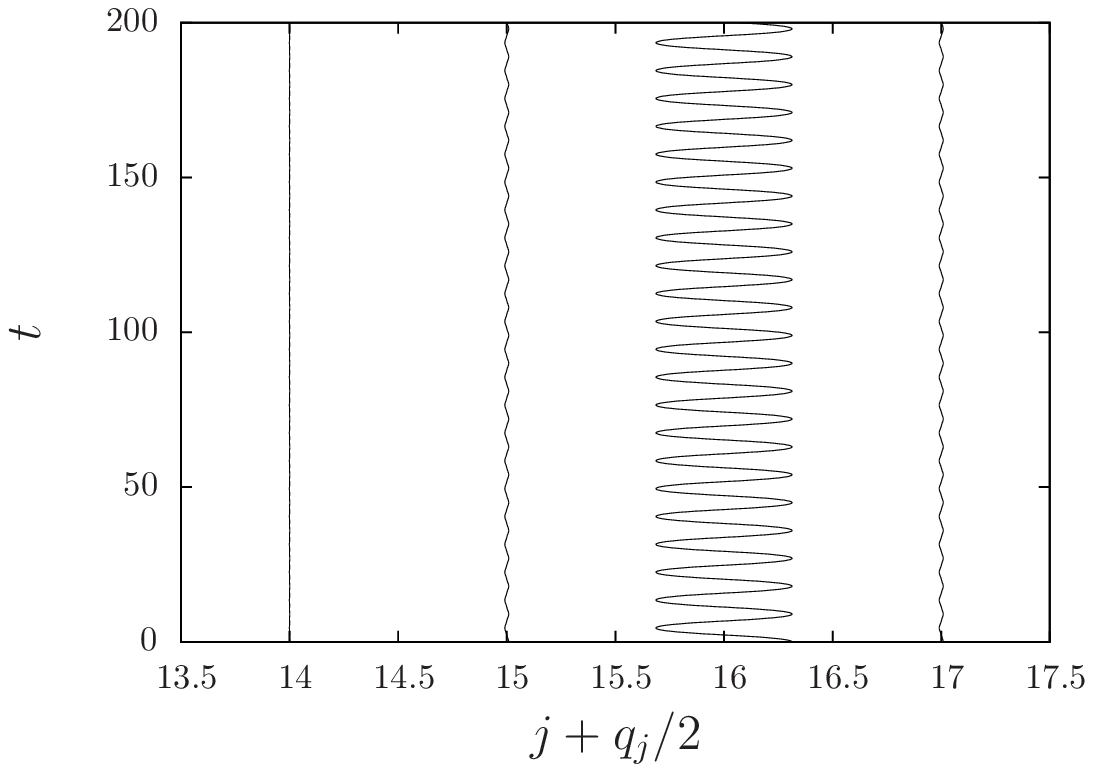}
  \includegraphics[width=3.5cm]{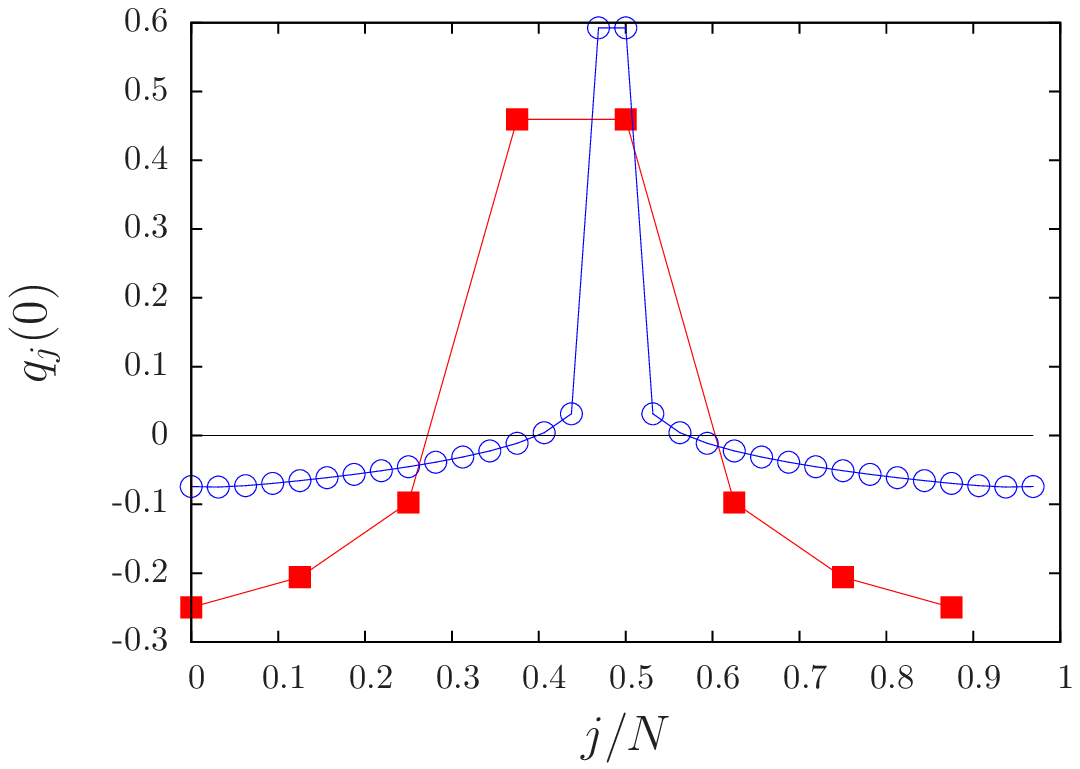}
  \includegraphics[width=3.5cm]{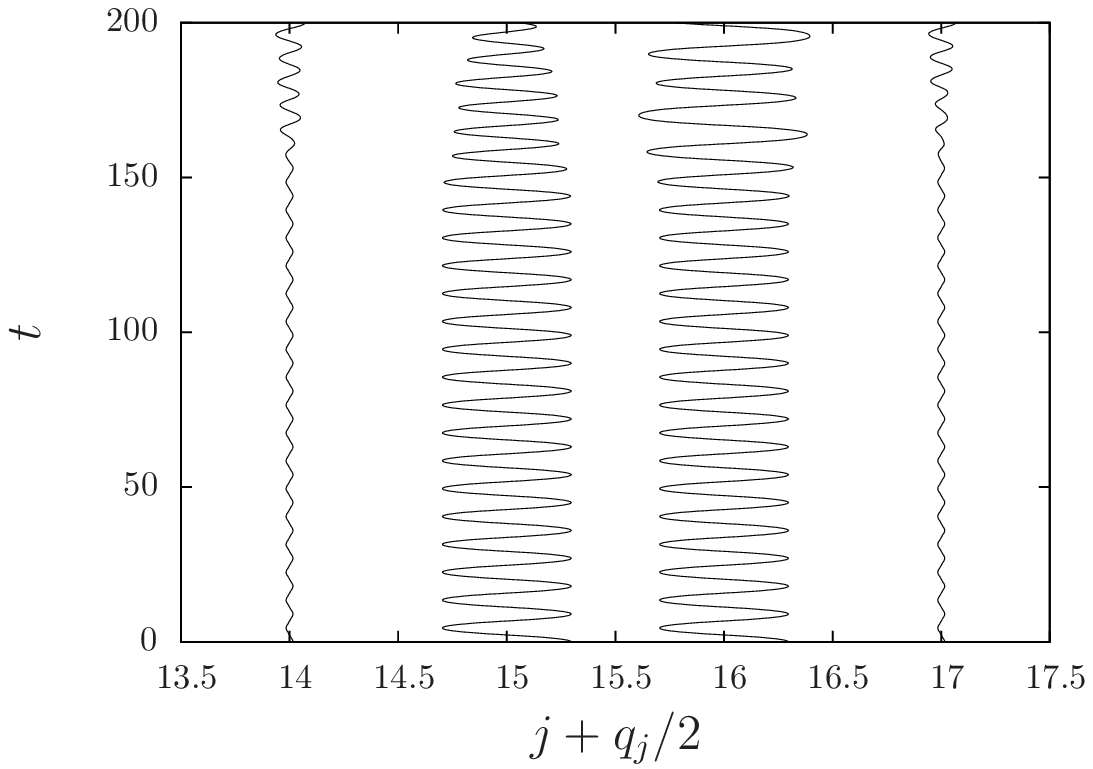}
  \includegraphics[width=3.5cm]{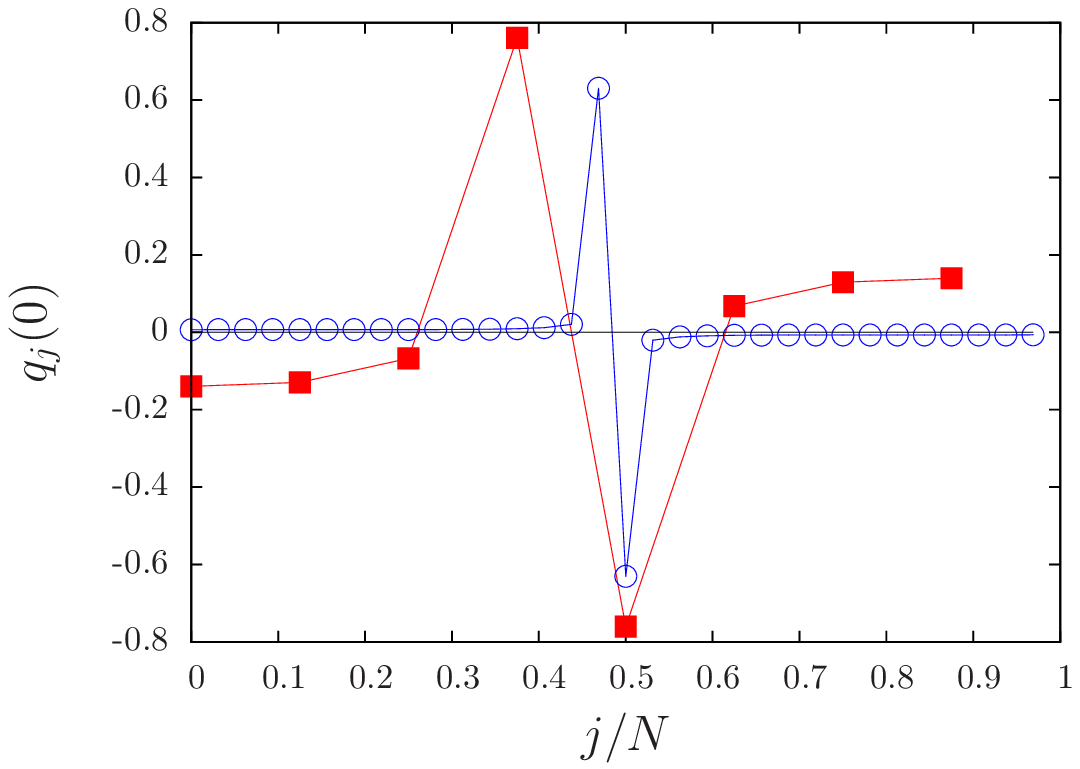}
  \includegraphics[width=3.5cm]{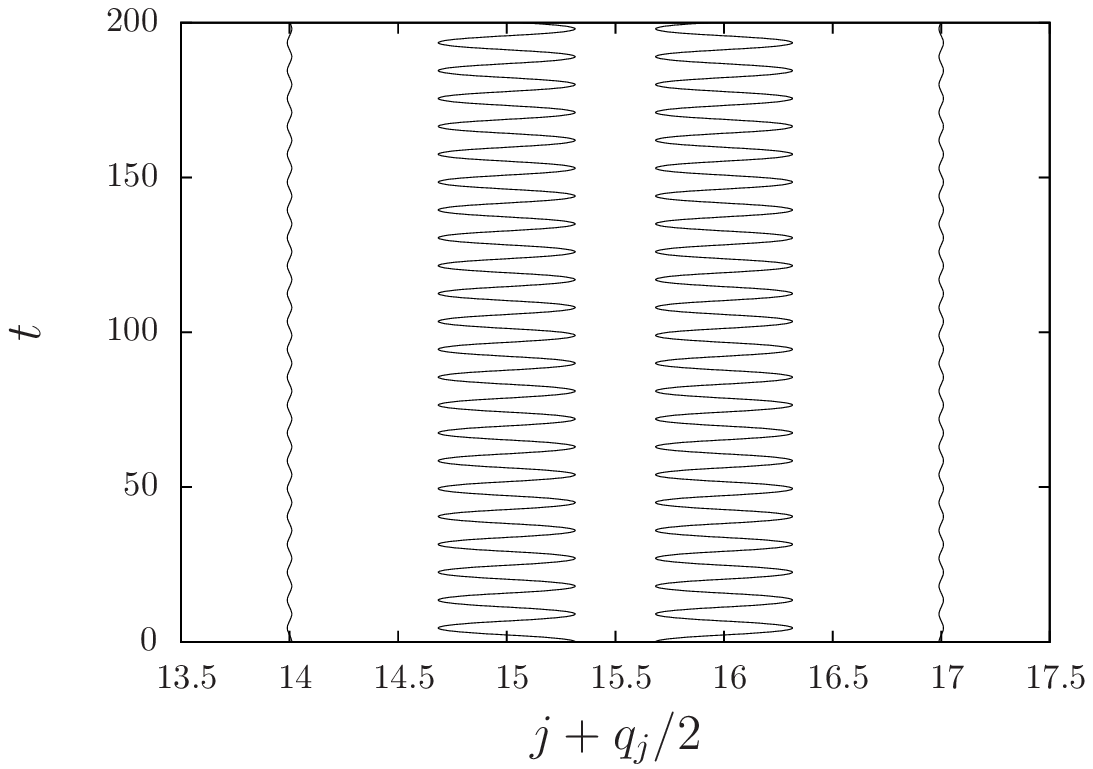}
  \caption{(color online)
    Same with Fig.\ref{fig:HMF} but in the $\alpha$-FPU-$\beta\gamma$ model
    with the fixed boundary condition. $\alpha=0.5$.
    The period is $T=9$ in all the panels.
    In right panels, the amplitudes are divided by $2$ for a graphical reason.
    }
  \label{fig:FPU}
\end{figure}

It is known that two types of DB solutions are possible: ST
modes\cite{Sievers1988} and P modes\cite{Page1990} in nonlinear lattices.
Periodic solutions of ST, P-in-phase and P-anti-phase modes
are exhibited in Fig.\ref{fig:HMF} for the $\alpha$-HMF model
and in Fig.\ref{fig:FPU} for the $\alpha$-FPU-$\beta\gamma$ model
with $\alpha=0.5$ in both.
The periods are $T=15$ for the $\alpha$-HMF model
and $T=9$ for the $\alpha$-FPU-$\beta\gamma$ model.
The corresponding frequencies are $\omega\simeq 0.42$
and $\omega\simeq 0.70$ respectively,
and they are in the band gap below
(see Figs. \ref{fig:EigenValuesPeriodic} and \ref{fig:EigenValuesFixed}).

It is worth commenting that, in the two long-range systems,
both in-phase and anti-phase modes are obtained in the band gap below.
This observation gives a sharp contrast with the
discrete nonlinear Klein-Gordon system,
which has the in-phase in the band gap below,
but has the anti-phase in the band gap above \cite{Campbell2004}.

\begin{table}
  \centering
  \begin{tabular}{l|ll|ll}
    \hline
    & \multicolumn{2}{|c|}{$\alpha$-HMF} & \multicolumn{2}{|c}{$\alpha$-FPU-$\beta\gamma$} \\
    & $N=8$ & $N=32$ & $N=8$ & $N=32$ \\
    \hline
    ST & 1.0293 & 1.0000 & 1.0000 & 1.0000 \\
    P-in & 195.57 & 557.04 & 5.4527 & 7.2066 \\
    P-anti & 503.98 & 151.27 & 1.5271 & 1.0004 \\
    \hline
  \end{tabular}
  \caption{Maximum absolute value of the Floquet exponents.}
  \label{tab:maxFloquet}
\end{table}

Stability of the periodic solutions are researched
by computing the maximum value of the Floquet exponents
arranged in Tab.\ref{tab:maxFloquet}.
The periodic solutions of P modes are unstable,
and the periodicity breaks due to small disturbance.
We remark that the stability of ST mode changes
from unstable to stable in the $\alpha$-HMF model.
This fact suggests that the stability depends on the number of particles $N$
in long-range systems.

{\it Summary and discussions:}
We have proposed a new mechanism, the long-range couplings,
to open a band gap below {\it without} on-site potential.
This mechanism is universal for any long-range systems
irrespective of details of the coupling functions.
Existence of periodic solutions are demonstrated in two systems,
each of which has the periodic boundary condition
and the fixed boundary condition.
Moreover, they have different topology of the inner degree of freedom:
the unit circle as XY spin and the real axis as the conventional spring system.

A band gap below opens even in $\alpha >1$
if numbers of particles is small. Thus, DB can be excited in small
numbers lattice systems
having all-to-all interaction with short range interaction of $\alpha>1$.

The results presented in this paper can be applied to various lattice
systems with long-range interactions such as DNA
models\cite{Cuevas2002} and spin
systems\cite{Lai1997,Schwarz1999,Zolotaryuk2001,Tang2014,Lakshmanan2014}.
It is also expected that another type of DB can be expected in MEMS
cantilever arrays\cite{English2012, English2013} by considering
long-range interactions between cantilevers,
through the overhang for instance.

It has to be done to consider traveling DBs in long-range systems.
As discussed above, $\alpha>1$ may give a band gap below
for small number of particles, and it might be interesting
if traveling DBs are realized with low frequencies under the concept of PISL, 
which has $\alpha=2$ for the FPU-$\beta$ type system.
Another important thing to do is to propose an experimental setting
to observe DBs being based on the mechanism proposed in this article.

Dynamics of long-range systems is described by the Vlasov equation
in the limit of large population \cite{Bachelard2011}.
The Vlasov equation has infinite number of the Casimir invariants,
and they provide strange critical exponents
\cite{Ogawa2014a,Ogawa2014b,Ogawa2015}
and finite-size fluctuation \cite{Yamaguchi2016}.
It has to reveal how the Casimirs work in DBs.

Finally, we stress that introducing long-range couplings is a quite new idea
to make DBs with low frequencies.
We expect that this idea creates a new bridge between material science
and nonlinear dynamical theory,
and induces successive researches.

\acknowledgements
YYY acknowledges the supports of
JSPS KAKENHI Grant Number 16K05472.
YD acknowledges the supports of
JSPS KAKENHI Grant Number 16K05041.


\begin{thebibliography}{99}
\bibitem{Sievers1988}
  A. J. Sievers and S. Takeno,
  Intrinsic Localized Modes in Anharmonic Crystals,
  Phys. Rev. Lett. {\bf 61}, 970 (1988).

\bibitem{Campbell2004}
  D. K. Campbell, S. Flach and Y. S. Kivshar,
  Localizing Energy Through Nonlinearity and Discreteness,
  Physics Today, {\bf 57}, 43 (2004).

\bibitem{Flach2008}
  S. Flach and A. V. Gorbach,
  Discrete breathers -- Advances in theory and applications,
  Phys. Rep. {\bf 467}, 1 (2008).

\bibitem{Yoshimura2015}
  K. Yoshimura, Y. Doi and M. Kimura,
  Localized Modes in Nonlinear Discrete Systems,
  in {\it Progress in Nanophotonics 3},
  edited by M. Ohtsu and T. Yatsui
  (Springer International Publishing, Cham, 2015)
  pp. 119-166.

\bibitem{Dmitriev2015}
  S. V. Dmitriev, A. P. Chetverikov and M. G. Velarde,
  Discrete breathers in 2D and 3D crystals,
  Physica status solidi (b) {\bf 252}, 1682 (2015).

\bibitem{Bajars2015}
  J. Bajars, J. C. Eilbeck and B. Leimkuhler,
  Numerical Simulations of Nonlinear Modes in Mica: Past, Present and Future,
  in {\it Quodons in Mica},
  edited by J. F. R. Archilla, N. Jim{\'e}nez, V. J. S{\'a}nchez-Morcillo and L. M. Garc{\'i}a-Raffi
  (Springer International Publishing, Cham, 2015)
  pp. 35-67.

\bibitem{Lai1997}
  R. Lai and A. J. Sievers,
  Intrinsic localized spin wave modes in easy-axis antiferromagnetic chains,
  J. Appl. Phys. {\bf 81}, 3972 (1997).

\bibitem{Schwarz1999}
  U. T. Schwarz, L. Q. English and A. J. Sievers,
  Experimental Generation and Observation of Intrinsic Localized Spin Wave Modes in an Antiferromagnet,
  Phys. Rev. Lett. {\bf 83}, 223 (1999).

\bibitem{Zolotaryuk2001}
  Y. Zolotaryuk, S. Flach and V. Fleurov,
  Discrete breathers in classical spin lattices,
  Phys. Rev. B {\bf 63}, 214422 (2001).

\bibitem{Tang2014}
  B. Tang, D.-J. Li and Y. Tang,
  Spin discrete breathers in two-dimensional square anisotropic ferromagnets,
  Physica Scripta {\bf 89}, 095208 (2014).

\bibitem{Lakshmanan2014}
  M. Lakshmanan, B. Subash and A. Saxena,
  Intrinsic localized modes of a classical discrete anisotropic Heisenberg ferromagnetic spin chain,
  Phys. Lett. A {\bf 378}, 1119 (2014).

\bibitem{English2012}
  L. Q. English, F. Palmero, P. Candiani, J. Cuevas, R. Carretero-Gonz\'alez, P. G. Kevrekidis and A. J. Sievers,
  Generation of Localized Modes in an Electrical Lattice Using Subharmonic Driving,
  Phys. Rev. Lett. {\bf 108}, 084101 (2012).

\bibitem{English2013}
  L. Q. English, F. Palmero, J. F. Stormes, J. Cuevas, R. Carretero-Gonz\'alez and P. G. Kevrekidis,
  Nonlinear localized modes in two-dimensional electrical lattices,
  Phys. Rev. E {\bf 88}, 022912 (2013).

\bibitem{Sato2003}
  M. Sato, B. E. Hubbard, A. J. Sievers, B. Ilic, D. A. Czaplewski and H. G. Craighead,
  Observation of Locked Intrinsic Localized Vibrational Modes in a Micromechanical Oscillator Array,
  Phys. Rev. Lett. {\bf 90}, 044102 (2003).

\bibitem{Kimura2009}
  M. Kimura and T. Hikihara,
  Capture and release of traveling intrinsic localized mode in coupled cantilever array,
  Chaos {\bf 109}, 013138 (2009).

\bibitem{Flach1998}
  S. Flach,
  Breathers on lattices with long range interaction,
  Phys. Rev. E {\bf 58}, R4116 (1998).

\bibitem{Cuevas2002}
  J. Cuevas, J. F. R. Archilla, Y. B. Gaididei and F. R. Romero,
  Moving breathers in a DNA model with competing short-and long-range dispersive interactions,
  Physica D {\bf 163}, 106 (2002).

\bibitem{Miloshevich2017}
  G. Miloshevich, J. P. Nguenang, T. Dauxois, R. Khomeriki and S. Ruffo,
  Traveling solitons in long-range oscillator chains,
  J. Phys. A {\bf 50}, 12LT02 (2017).

\bibitem{Doi2016}
  Y. Doi and K. Yoshimura,
  Symmetric Potential Lattice and Smooth Propagation of Tail-Free Discrete Breathers,
  Phys. Rev. Lett. {\bf 117}, 014101 (2016).

\bibitem{Campa2009}
  A. Campa, T. Dauxois and S. Ruffo,
  Statistical mechanics and dynamics of solvable models with long-range interactions,
  Phys. Rep. {\bf 480}, 57 (2009).

\bibitem{Page1990}
  J. B. Page,
  Asymptotic solutions for localized vibrational modes in strongly anharmonic periodic systems,
  Phys. Rev. B {\bf 41}, 7835 (1990).

\bibitem{Bachelard2011}
  R. Bachelard, T. Dauxois, G. De Ninno, S. Ruffo and F. Staniscia,
  Vlasov equation for long-range interactions on a lattice,
  Phys. Rev. E {\bf 83}, 061132 (2011).

\bibitem{Ogawa2014a}
  S. Ogawa, A. Patelli and Y. Y. Yamaguchi,
  Non-mean-field critical exponent in a mean-field model: Dynamics versus statistical mechanics,
  Phys. Rev. E {\bf 89}, 032131 (2014).

\bibitem{Ogawa2014b}
  S. Ogawa and Y. Y. Yamaguchi,
  Nonlinear response for external field and perturbation in the Vlasov system,
  Phys. Rev. E {\bf 89}, 052114 (2014).

\bibitem{Ogawa2015}
  S. Ogawa and Y. Y. Yamaguchi,
  Landau-like theory for universality of critical exponents in quasistationary states of isolated mean-field systems,
  Phys. Rev. E {\bf 91}, 062108 (2015).

\bibitem{Yamaguchi2016}
  Y. Y. Yamaguchi,
  Strange scaling and relaxation of finite-size fluctuation in thermal equilibrium,
  Phys. Rev. E {\bf 94}, 012133 (2016).

\end{thebibliography}
\end{document}